\documentstyle[12pt]{article}

\def\+{{+\!\!\!+}}
\def\pp{\mbox{\tiny${}_{\stackrel\+ =}$}}

\def\d{\partial}

\def\th{\theta}

\def\P{\Phi}

\def\e{\varepsilon}

\def\pmb#1{\setbox0=\hbox{#1}%
\kern.0em\copy0\kern-\wd0
\kern-.04em\copy0\kern-\wd0
\kern.08em\copy0\kern-\wd0
\kern-.04em\raise.0433em\box0 }         

\def\rank{\textstyle{\rm{rank}}}

\newcommand{\nc}{\newcommand}
\nc{\beq}{\begin{equation}}
\nc{\eeq}[1]{\label{#1}\end{equation}}
\nc{\ber}{\begin{eqnarray}}
\nc{\eer}[1]{\label{#1}\end{eqnarray}}
\nc{\pek}[1]{\cite{#1}}
\nc{\enr}[1]{(\ref{#1})}
\nc{\kal}[1]{{\cal{#1}}}
\nc{\dott}{\;\cdot\;}

\begin{document}
\newcommand{\inv}[1]{{#1}^{-1}} 
\renewcommand{\theequation}{\thesection.\arabic{equation}}
\newcommand{\be}{\begin{equation}}
\newcommand{\ee}{\end{equation}}
\newcommand{\bea}{\begin{eqnarray}}
\newcommand{\eea}{\end{eqnarray}}
\newcommand{\re}[1]{(\ref{#1})}
\newcommand{\qv}{\quad ,}
\newcommand{\qp}{\quad .}
\begin{center}

                                \hfill   hep-th/0210043\\
\vskip .3in \noindent

\vskip .1in

{\large \bf{Poisson geometry of sigma models with extended supersymmetry}}
\vskip .2in

{\bf Simon Lyakhovich}$^a$\footnote{e-mail address: sll@phys.tsu.ru}
 and  {\bf Maxim Zabzine}$^{b}$\footnote{e-mail address: zabzine@fi.infn.it} \\

\vskip .15in

\vskip .15in
$^a${\em  Department of Theoretical Physics, Tomsk State University, \\
 634050, Lenin av. 36, Tomsk, Russia}\\
\vskip .15in
$^b${\em INFN Sezione di Firenze, Dipartimento di Fisica, \\
 Via Sansone 1, I-50019 Sesto F.no (FI), Italy}

\bigskip

\vskip .1in

\end{center}
\vskip .4in

\begin{center} {\bf ABSTRACT }
\end{center}
\begin{quotation}\noindent
We consider a general N=(2,2) non-linear sigma model with a torsion.
 We show that the consistency of N=(2,2)
 supersymmetry implies that the target manifold is necessary equipped with
 two (in general, different) Poisson structures.
 Finally we argue that the Poisson
 geometry of the target space is a characteristic feature of the
 sigma models with extended supersymmetry.
 \end{quotation}
 \vfill
 \eject

\section{Introduction}

 The supersymmetric non-linear sigma models with extended supersymmetry
  play a prominent role both in physics and mathematics.
  From the physical point of view these models are of manifest interest
 to the string theory. On the mathematical side there is
 an intriguing relation between these models and the complex geometry
 which has been well recognized already in eighties
 \cite{Zumino:1979et}-\cite{Hitchin:1986ea}.

 In the
 present paper we would like to take a fresh look at the geometry of the
 general N=(2,2) supersymmetric two dimensional non-linear sigma model.
 Our study reveals a deep relationship between  Poisson
 geometry and supersymmetry. In particular we show that the consistency of N=(2,2)
 supersymmetry implies that the target manifold is necessary equipped with
 two (in general, different) Poisson structures.

 In a certain sense this situation is similar to the case
 of topological Poisson sigma models \cite{Ikeda:1993fh,SchStrobl} where the
 consistency of gauge symmetry requires the target manifold to have a Poisson
 structure. These models have been efficiently
 exploited for the deformation quantization and have brought a natural generating
 procedure \cite{CF1} for the Kontsevich star product
 \cite{Kontsevich}.
 The method of topological Poisson sigma models combined with the
 Fedosov approach \cite{Fedosov} allows to construct explicitly the
 covariant star-products \cite{CFT}.

Starting from Zumino's observation \cite{Zumino:1979et}
 that the target manifold
 of the N=2 supersymmetric sigma model
 needs to be K\"ahler, it has been recognized that supersymmetry and complex geometry
 go hand in hand.
 In \cite{Alvarez-Gaume:hm}, Alvarez-Gaum\'e and Freedman classified possible supersymmetric sigma models
 in two dimensions.
 Later in \cite{Gates:1983py} it was remarked that more general supersymmetric sigma model can be
 constructed by including Wess-Zumino term in the action.
  Gates, Hull and Ro\v{c}ek described
 the general N=2 supersymmetric sigma model with torsion \cite{Gates:1984nk}.
  The target manifold of such model should be a bihermitian complex manifold with
  extra non-trivial constraints on the complex structures. Ever since the work of Gates-Hull-Ro\v{c}ek
  these models have attracted a lot of attention, e.g. \cite{Buscher:uw}-\cite{Bogaerts:1999jc} (also
 the N=2 WZW models provide  important
 examples of this geometry, \cite{Spindel:1988nh}-\cite{Sevrin:1988ps}).
 However the nature of those
 non-trivial constraints on the complex structure remains unclear.
 In this paper we
 reformulate and interpret these non-trivial constraints on the complex structures.
  More specifically, we study these constraints on the background
  geometry not in the terms of complex structures and exterior forms, but in the 
  terms of the antisymmetric contravariant tensor analysis (using the Schouten
  brackets). As a result  we observe that the consistency conditions of
  the N=(2,2) supersymmetry for the model require a bi-Poisson
  structure to exist on the target manifold.

This paper is organized as follows. In Section~\ref{s:susy} we review
the results from the Gates-Hull-Ro\v{c}ek work \cite{Gates:1984nk} on
 the N=2 supersymmetric sigma model with torsion. Following this
 introductory section in Section~\ref{s:Poisson} we proceed to the
 reformulation of the geometry in terms of contravariant antisymmetric
  tensors and the Schouten bracket. Due to this reformulation of the
  geometric conditions we arrive at the claim that the bihermitian
 complex manifold which supports the N=(2,2) model is bi-Poisson (i.e.,
 there exist  two Poisson structures).
 In next Section~\ref{s:geometry} using the fact of the existence of
 two Poisson structures we explore the geometry of the manifolds and
 propose a classification program for them.  In
 Section~\ref{s:generalizations} we look at the generalizations of the
 presented logic for the N=4 supersymmetric sigma model.  Finally in
 Section~\ref{s:discussion} we give a summary of the paper and discuss
 the open problems. We argue that the appearance of the Poisson
 structures is a common feature of the sigma models with extended
 supersymmetry.

 In this paper we attempt to clarify the relation of supersymmetry and
 Poisson geometry in terms accessible to both mathematicians
 and physicists. To this end, we review some basic and well known notions in terms
 intended to make them accessible to a new audience.

\section{N=(2,2) supersymmetric sigma model}
\label{s:susy}

In this section we review the results on the general N=(2,2) supersymmetric sigma model
 from the original work \cite{Gates:1984nk}.
 This will allow us to introduce the notation and some relevant concepts.

 Let us start from the general N=1 sigma model which is written in N=1 superfields
 (see Appendix for the conventions)
\beq
 S= \int d^2\sigma\,d^2\theta\,\,D_+\Phi^\mu D_- \Phi^\nu (g_{\mu\nu}(\Phi)
 + B_{\mu\nu}(\Phi)) .
\eeq{actionB}
The action (\ref{actionB}) is manifestly supersymmetric under the supersymmetry
 transformations
\beq
 \delta^1(\epsilon)  \Phi^\mu  = i (\epsilon^+ Q_+ + \epsilon^- Q_-) \Phi^\mu ,
\eeq{susyI}
 which form the standard supersymmetry algebra
\beq
 [ \delta^1(\epsilon_1), \delta^1(\epsilon_2)] \Phi^\mu =
 2i \epsilon_1^+ \epsilon_2^+ \d_\+ \Phi^\mu
 + 2i \epsilon_1^- \epsilon_2^- \d_= \Phi^\mu .
\eeq{susyIalgebra}
 We look for additional supersymmetry transformations of the form
$$ \delta^2(\epsilon) \Phi^\mu = \epsilon^\alpha D_\alpha \Phi^\nu L^\mu_{\,\,\nu}(\Phi) + \epsilon^\alpha
 (\gamma_3 D)_\alpha \Phi^\nu M^\mu_{\,\,\nu}(\Phi) = $$
\beq
 =\epsilon^+ D_+ \Phi^\nu J^\mu_{+\nu}(\Phi)
  + \epsilon^- D_- \Phi^\nu J^\mu_{-\nu}(\Phi) ,
\eeq{secsupfl}
 where
\beq
 L^\mu_{\,\,\nu} =\frac{1}{2} (J^\mu_{+\nu} + J^\mu_{-\nu}),\,\,\,\,\,\,\,\,\,\,\,\,\,\,
  M^\mu_{\,\,\nu} =\frac{1}{2} (J^\mu_{+\nu} - J^\mu_{-\nu}) .
\eeq{definLM}
 Classically the ansatz (\ref{secsupfl}) is unique for dimensional reasons.
It turns out that the action (\ref{actionB}) is invariant under the transformations
 (\ref{secsupfl}) provided that
\beq
 J_{\pm\rho}^\mu g_{\mu\nu} = - g_{\mu\rho} J^\mu_{\pm\nu}
\eeq{bla}
 and
\beq
 \nabla^{(\pm)}_\rho J^\mu_{\pm\nu} \equiv J^\mu_{\pm\nu,\rho} +
 \Gamma^{\pm\mu}_{\,\,\rho\sigma} J^\sigma_{\pm\nu} - \Gamma^{\pm\sigma}_{\,\,\rho\nu}
 J^\mu_{\pm\sigma}=0
\eeq{nablJH}
 where one defines two affine connections
\beq
 \Gamma^{\pm\mu}_{\,\,\rho\nu} = \Gamma^{\mu}_{\,\,\rho\nu} \pm g^{\mu\sigma} H_{\sigma\rho\nu}
\eeq{defaffcon}
with $H$ being the torsion three form,
\beq
H_{\mu\rho\sigma} = \frac{1}{2}
 (B_{\mu\rho,\sigma} + B_{\rho\sigma,\mu} + B_{\sigma\mu,\rho}).
\eeq{Hdefin}
  Next  we  have to require the standard on-shell N=2 supersymmetry algebra, i.e. the first
 supersymmetry transformations (\ref{susyI}) and the second supersymmetry transformations (\ref{secsupfl})
 commute
\beq
 [\delta^2(\epsilon_1), \delta^1(\epsilon_2)]\Phi^\mu = 0
\eeq{commtwosusy}
and the commutator of two second supersymmetry transformations
 $$ [\delta^2(\epsilon_1), \delta^2(\epsilon_2)]\Phi^\mu = -2i \epsilon_1^+ \epsilon_2^+ \d_\+ \Phi^\lambda
 (J^\nu_{+\lambda} J^\mu_{+\nu}) -2i \epsilon_1^- \epsilon_2^- \d_=  \Phi^\lambda
 (J^\nu_{-\lambda} J^\mu_{-\nu}) + $$
\beq
+ \epsilon_1^+ \epsilon_2^+ D_+\Phi^\lambda D_+\Phi^\rho
 {\cal N}^\mu_{\,\,\lambda\rho}(J_+) +  \epsilon_1^- \epsilon_2^- D_-\Phi^\lambda D_-\Phi^\rho
 {\cal N}^\mu_{\,\,\lambda\rho}(J_-)
\eeq{secondsusy}
 should satisfy the same algebra as the first (\ref{susyIalgebra}), i.e.
\beq
[\delta^2(\epsilon_1), \delta^2(\epsilon_2)]\Phi^\mu =2i \epsilon_1^+ \epsilon_2^+ \d_\+ \Phi^\mu
 +  2i \epsilon_1^- \epsilon_2^- \d_=  \Phi^\mu .
\eeq{secondlikefirst}
 In (\ref{secondsusy})
 ${\cal N}$ is the Nijenhuis tensor which is defined as follows
\beq
 {\cal N}^{\rho}_{\,\,\mu\nu} (J_{\pm}) = J^\gamma_{\pm\mu} \d_{[\gamma} J^\rho_{\pm\nu]}
 - J^\gamma_{\pm\nu} \d_{[\gamma} J^\rho_{\pm\mu]} .
\eeq{intergpl11}
 In order to arrive at the expression (\ref{secondsusy}) we have used  the property (\ref{nablJH})
 and the equations of motion
\beq
 D_+ D_- \Phi^\mu + \Gamma^{-\mu}_{\,\,\rho\sigma} D_+\Phi^\rho D_-\Phi^\sigma = 0 ,
\eeq{eqsmot}
 which  follow from the action (\ref{actionB}).
 If we want the algebra (\ref{secondsusy}) to be identical to the algebra
 (\ref{susyIalgebra}) we have to require
 the following properties of $J_{\pm}$
\beq
 J^\mu_{\pm\nu} J^\rho_{\pm\mu} = -\delta ^\rho_{\,\,\nu},\,\,\,\,\,\,\,\,\,\,\,\,\,\,\,\,\,\,
\eeq{complstruct}
\beq
 {\cal N}^{\rho}_{\,\,\mu\nu} (J_{\pm}) = 0
\eeq{intergpl}
  Thus the supersymmetry algebra requires
 that $J_{\pm}$ correspond to two complex structures.

 This is the full description of the most general N=(2,2) sigma model. Thus the target
 manifold should be a bihermitian complex manifold (i.e., there are two complex structures
 and a metric is Hermitian with respect to both) and the two complex structures should
 be covariantly constant, with respect to the different connections however.
 The main question we would like to address is under what conditions on two
 complex structures (or their K\"ahler forms) one can construct the appropriate
 affine connections such that (\ref{nablJH}) is fulfilled.

 Before answering this question let us explore the properties of the
 target manifold.
 The vanishing of the
 Nijenhuis tensor (\ref{intergpl11}) and the condition (\ref{nablJH}) imply that the
 complex structures should preserve the torsion
 \beq
 H_{\delta\nu\lambda} = J^\sigma_{\pm\delta} J^\rho_{\pm\nu} H_{\sigma\rho\lambda} +
 J^\sigma_{\pm\lambda} J^\rho_{\pm\delta} H_{\sigma\rho\nu}+
 J^\sigma_{\pm\nu} J^\rho_{\pm\lambda} H_{\sigma\rho\delta} .
\eeq{inegrabtroB}
 Introducing the K\"ahler forms
\begin{equation}
 \omega_{\pm}\equiv gJ_{\pm}
  \label{og}
  \end{equation}
   and  using equations (\ref{nablJH}) one can find $d\omega_{\pm}$
\beq
(d\omega_\pm)_{\rho\mu\nu} =  \pm \left ( H_{\sigma\rho\mu} J^\sigma_{\pm\nu} +
 H_{\sigma\mu\nu} J^\sigma_{\pm\rho} + H_{\sigma\nu\rho} J^\sigma_{\pm\mu}\right ) ,
\eeq{definitdJ}
where we use the following normalization to the exterior differential
\beq
(d\omega_\pm)_{\lambda\sigma\gamma} = \frac{1}{2}(\d_\lambda \omega_{\pm\sigma\gamma} +
\d_\sigma \omega_{\pm\gamma\lambda} + \d_\gamma \omega_{\pm\lambda\sigma}) .
\eeq{defextderJpl}
 Now by combining  (\ref{inegrabtroB}) and (\ref{definitdJ}) one arrives at the following
 property
\beq
 H_{\mu\nu\rho} = - J^\lambda_{+\mu} J^\sigma_{+\nu} J^\gamma_{+\rho} (d \omega_+)_{\lambda\sigma\gamma} =
 J^\lambda_{-\mu} J^\sigma_{-\nu} J^\gamma_{-\rho} (d\omega_-)_{\lambda\sigma\gamma}  .
\eeq{HtermJplm}
 The first equality in expression (\ref{HtermJplm})
 can be considered as a definition
 of the torsion. Starting from the complex structure, let say $J_{+}$, one can construct
 the torsion (\ref{HtermJplm}) such that $\nabla^{(+)}_\rho J^\mu_{+\nu} =0$ where $\nabla^{(+)}$
 is defined in (\ref{nablJH}), \cite{Yano1}.
 Thus the only non-trivial constraint which arises from (\ref{HtermJplm})
  is the relation between $(J_{+}, \omega_+)$ and $(J_{-}, \omega_-)$.
  However, in the present form the relation (\ref{HtermJplm})
 tells us very little about the geometry of the target manifold.

 To summarize the discussion, let us state the geometric
 data which gives the  necessary and sufficient conditions for the 
 target manifold to admit the N=(2,2) sigma model.
   The target manifold ${\cal M} (g, J_\pm)$ 
 should be equipped with two complex structures $J_{\pm}$ and the
 metric $g$ which is Hermitian with respect to both complex structures $J_\pm$.
   In addition, the complex structures $J_{\pm}$ together with
 their K\"ahler forms $\omega_{\pm}$ (\ref{og}) should satisfy the relation
 (\ref{HtermJplm}), defining the torsion $H$.

There is another property which arises in the construction, the
closedness of torsion form, $dH=0$.  However this property is only
related to the fact that the equations of motion are derived from the
Lagrangian, which may involve the closed torsion only.
 At the level of symmetries and
classical dynamics, this property is not required (i.e., the
field equations and the N=2 supersymmetry remain consistent even when $dH \neq 0$
in the model)
and therefore in our further discussion we treat the closedness of the
torsion as a secondary property.

\section{Poisson structures of (2,2) sigma model}
\label{s:Poisson}

In this
 section we reformulate some of the previous conditions in new terms.
 In particular we give a new form to the constraint
(\ref{HtermJplm}) relating two complex structures involved in the
second supersymmetry transformation, so that
it reveals the Poisson structures on the target manifold of the N=(2,2)
 sigma model.

It turns out that it is appropriate to describe the target manifold geometry
in terms of {\it contravariant} antisymmetric tensor fields instead
of {\it covariant} ones like metric $g$, exterior forms $\omega_\pm$
(\ref{og}) and their differentials,
as it has been done in previous section.
Before doing this, let us remind
the relevant notions related to the differential
calculus of contravariant tensor fields.

Given two antisymmetric contravariant tensor fields (multivectors)
$\alpha$ and
$\beta$ of the rank $|\alpha| $ and $|\beta|$ respectively
 one may define a Schouten bracket \cite{Schouten} of
them being a tensor of rank $|\alpha|+|\beta|-1$:
$$
\alpha =
 \alpha^{i_1 \ldots i_{|\alpha|}} \partial_{i_1}\wedge \ldots \wedge
\partial_{i_{|\alpha|}} \, , \, \, \beta = \beta^{i_1 \ldots i_{|\beta|}}
\partial_{i_1}\wedge \ldots \wedge \partial{i_{|\beta|}} \, ,
$$
\beq
[\alpha, \beta]_s \equiv ( |\beta| \partial_j
\alpha^{i_1 \ldots i_{|\alpha|}}
\beta^{j i_{|\alpha|+1} \ldots i_{|\beta|+|\alpha|-1}} -
 |\alpha| \alpha^{i_1 \ldots i_{|\alpha|-1} j} \partial_j
 \beta^{i_{|\alpha|} \ldots i_{|\beta|+|\alpha|-1}})
\partial_{i_1} \wedge \ldots \wedge
\partial_{i_{|\alpha| +|\beta| - 1}} .
 \label{defS}
\end{equation}
The Schouten
 bracket can be thought of as an extension of the Lie bracket of
 the vector fields to the case of multivectors.  Being defined
 for inhomogeneous tensor fields (by linearity), this bracket can be
 identified to the Batalin-Vilkovisky antibracket \cite{BV} on the odd cotangent
 bundle $\Pi T^* {\cal M}$ of the target manifold ${\cal M}$.
 The bracket (\ref{defS}) has  properties of the
 {\it odd} Poisson bracket (or anti-bracket that is the same)
 which is
 well-known in physics from the Batalin-Vilkovisky field-antifield formalism:
 (graded) symmetry, (graded) Jacobi identity and (graded) Liebnitz rule:
 \ber
 [\alpha, \beta]_s & = & - (-1)^{(|\alpha|-1)(|\beta|-1)}
  [\beta,  \alpha]_s \\
  \, [ \alpha , [\beta, \gamma]_s ]_s & = & [[\alpha , \beta]_s ,
 \gamma]_s + (-1)^{(|\alpha| -1)(|\beta|-1)}[\beta, [\alpha , \gamma ]_s ]_s
 \\ \, [\alpha, \beta \wedge \gamma ]_s & = & [\alpha , \beta ]_s \wedge
\gamma + (-1)^{(|\alpha|-1)|\beta|} \beta \wedge [\alpha, \gamma ]_s
\eer{propSchout}
Notice that the Schouten bracket of two even rank tensors is symmetric,
so the Schouten nilpotency condition is not automatically satisfied
for an even rank tensor. If second rank tensor $p$
 is Schouten-nilpotent on $\cal M$,
\beq
p=p^{\mu\nu}\partial_\mu \wedge \partial_\nu \, , \quad
[p,p]_s = 0
\eeq{nilp}
it defines the conventional (even) Poisson bracket
\beq
 \{ f, g \} \equiv p(df, dg) = p^{\mu\nu} \d_\mu f \, \d_\nu g,\,\,\,\,\,\,\,\,\,\,\,f(x), g(x)
 \in C^\infty({\cal M}),
\eeq{poissbr}
which is a bilinear map $C^\infty({\cal M})\times C^\infty({\cal M})
 \rightarrow C^\infty({\cal M})$. Because of (\ref{nilp}) the
 Poisson bracket (\ref{poissbr}) has ordinary antisymmetry property
 and satisfies standard Leibnitz rule and Jacobi identity.

The tensor $p$ satisfying (\ref{nilp})  is called a Poisson tensor
and the manifold  which admits such $p$ is called a Poisson manifold
(for a review of the Poisson geometry see, e.g. \cite{vaisman}).
 Notice that the Poisson geometry does not imply that
$p^{\mu\nu}$ has an inverse, moreover the rank of $p$ may be varying from point
to point on the manifold. In the case of nondegenerate
Poisson tensor we have a symplectic manifold.

Two Poisson tensors $p_1$ and $p_2$ are called compatible \cite{magri}
if they Schouten commute (i.e., $[p_1,p_2]_s=0$). Apparently any linear combination of two
compatible Poisson tensors $p_a, \, a=1,2$ is a Poisson tensor again. So if
the manifold manifold is equipped with two compatible Poisson structures
there exists a continuos family of the Poisson brackets
$p(k)= k^a p_a$, $k=(k^1,k^2) \in RP^1$ with any two elements
$p(k), p(l), \, \forall k,l \in RP^1$ compatible
to each other. The compatibility property
was first studied for Poisson structures in relation to
the integrable systems \cite{magri}.

Now let us reformulate the constraint (\ref{HtermJplm}) in the new terms.
Introduce the  contravariant tensors $j_{\pm}$ which are dual
to two-forms $\omega_{\pm}$ (\ref{og})
\beq
 j_{+}^{\mu\nu} \omega_{+\nu\rho}=\delta^\mu_{\,\,\rho},\,\,\,\,\,\,\,\,\,\,\,\,\,
 j_{-}^{\mu\nu} \omega_{-\nu\rho}=\delta^\mu_{\,\,\rho} .
\eeq{defcont}
Using the metric, we raise the indices in
(\ref{HtermJplm}) and rewrite this condition in the
 equivalent form in terms of $j_\pm$:
\beq
 2h = - [j_+, j_+]_s = [j_-, j_-]_s
\eeq{SchbrJJJ}
 where
 \beq
 h^{\mu\nu\rho} = g^{\mu\lambda} g^{\nu\sigma} g^{\rho\gamma}
 H_{\lambda\sigma\gamma} .
 \eeq{h}
 Just in terms of $j_{\pm}$ the condition (\ref{SchbrJJJ}) has the form
\beq
[j_+, j_+]_s + [j_-, j_-]_s = 0 .
\eeq{condnnn}
Thus for the  complex structures $J_{\pm}$
 on the bihermitian manifold, the condition (\ref{SchbrJJJ}) implies that
 there exists an affine connection such that the conditions (\ref{nablJH})
 are satisfied.
 Using the fact that both connections respect the
 metric (i.e., $\nabla^{(\pm)}_\rho g_{\mu\nu} =0$),
 one can show that (\ref{nablJH}) implies
 $j_-$ should Schouten commute to $j_+$
 \beq [j_-, j_+ ]_s = 0 .
 \eeq{fromcom}
 Introducing  the linear combinations of $j_{\pm}$
\beq
 l= \frac{1}{2}(j_+ + j_-),\,\,\,\,\,\,\,\,\,\,\,\,\,\,\,\,\,
 m= \frac{1}{2}(j_+ - j_-)
\eeq{lincombml}
the algebra (\ref{SchbrJJJ}) and (\ref{fromcom})
 can be equivalently rewritten as follows
\beq
  [l, l ]_s = 0,\,\,\,\,\,\,\,\,\,\,\,\,\,\,\,\,\,\,\,
  [m, m]_s = 0,\,\,\,\,\,\,\,\,\,\,\,\,\,\,\,\,\,\,\,
 [l, m]_s = - h .
\eeq{schoutnilp}
 Thus $l$ and $m$ are Schouten-nilpotent\footnote{Unlike
 $j_{\pm}$, $l$ and $m$ may be degenerate in a generic situation, for the details see
 the next section.}.
 Notice that the relations  (\ref{schoutnilp})
 are consistent because of the Schouten bracket properties
 (\ref{propSchout}) only, without requiring the torsion three-form $H$
 entering 3-vector $h$ to be closed.

 As a result of the previous discussion we see that $l$ and $m$ are Poisson structures
 which are not compatible however.
 Therefore the resulting manifold is a bi-Poisson (i.e., there are two
 Poisson structures, not an infinite family of the Poisson tensors).
 Let us now summarize this result.
 The manifold with two complex structures $J_+ , J_-$ and
 with the metric $g$ which is hermitian with respect to both structures
 will  admit the on-shell algebra (\ref{susyIalgebra}), (\ref{commtwosusy})
 and (\ref{secondlikefirst}) if and only if $(J_{+} \pm J_{-})g^{-1}$ are
 Poisson structures. We have already proved this statement in one
 direction: starting from the bihermitian manifold and using the
 specific form of the covariant constancy (\ref{nablJH}) of the complex
 structures, we have shown that the algebra (\ref{schoutnilp}) is
 satisfied. In opposite direction, having the bihermitian manifold with
 the algebra $[l,l]_s=0$ and $[m,m]_s=0$ one immediately arrives at the
 property (\ref{condnnn}) which is equivalent to the property
 (\ref{HtermJplm}) which defines the torsion such that (\ref{nablJH})
 is satisfied.  Therefore the on-shell algebra\footnote{On-shell
 algebra means that we are allowed to use the equations of motion while
 calculating the algebra. The problem of existence of an off-shell
 realization of the same algebra is a different problem
 which is not discussed here.}
 (\ref{susyIalgebra}), (\ref{commtwosusy}) and (\ref{secondlikefirst})
  is fulfilled. In order to have a Lagrangian formulation of the model
 one has to require that $dH=0$ which in fact leads to the extra
 requirement on the relation between the Poisson structures $l$ and $m$. However
 it does not affect their existence or non existence.

To summarize the discussion, the target manifold
$\cal M$ of the N=(2,2) sigma model can be completely characterized by
the following geometric data:
 $\cal M$ should be equipped with two complex structures
 $J_{\pm}$ and the  metric $g$ which is Hermitian with respect
 to both complex structures.
 In addition
 $(J_{+} + J_{-})g^{-1}$ and $(J_{+} - J_{-})g^{-1}$ should be the
 Poisson bi-vectors. The torsion $H$ is defined through the Schouten bracket
 of these two Poisson structures.

We would like to stress the fact that when either $l$ or $m$ vanishes
identically on $\cal M$ we end up with the standard K\"ahler case and
the well-known realization of N=2 supersymmetry algebra.  Therefore the
K\"ahler case is automatically included in the above discussion and
 corresponds to the ``singular'' case when either $J_+=J_-$ or
 $J_+=-J_-$.

\section{Poisson geometry of the target manifold: \\
local analysis and special cases} \label{s:geometry}

In this section we would like to analyse the immediate consequences of
the fact that the manifold is bi-Poisson. We  explore the
 local geometry of this manifold.  In particular we study
 the structure and intersections of the symplectic leaves of two
 Poisson structures defined by bi-vectors $l$ and $m$.

 Also we discuss in more details some special
 cases  of this geometry and propose a program for the classification of
 this type of geometries.

 We start by recalling the picture where one can
 (locally) think of a Poisson manifold
 as a union of symplectic leaves fitting together in a smooth way.
 Let ${\cal M}$ be a Poisson manifold
 with the Poisson structure $p^{\mu\nu}$, $\mu,\nu =1,2,...,d$ and let us choose the point
 $x_0$ such that in its neighbourhood $\rank(p)$
 is constant\footnote{Such a
 point is called a regular.
 On a non-regular Poisson manifold there are
 singular points which do not have this property \cite{vaisman}.
  We do not discuss here these points and their neighbourhoods.}.
  Contraction of the bi-vector $p$ to any
  one-form $e$, being a section of $T^*{\cal M}$, \,
  $e \not \in  {\rm ker} (p) $,
  provides a vector field which is called a locally Hamiltonian,
  if locally $e=df$ for some function $f \in C^\infty ({\cal M})$.
  So we have a set of
  locally Hamiltonian vector fields \beq v^\mu_k = p^{\mu\nu} \d_\nu
 f_k,\,\,\,\,\,\,\,\,\,\,\,\,\,\,k=1,2,...,n, n= d-\rank(p(x_0))
 \eeq{setHMF}
 which are in involution, i.e. the Lie bracket of two locally
 Hamiltonian vector fields is  a locally Hamiltonian field again
  \beq \{ v_i,
 v_j\}^\mu_{Lie} = p^{\mu\rho} \d_\rho \left ( (\d_\nu f_j) p^{\nu\lambda}
 (\d_\lambda f_i)\right ) .  \eeq{LiebrHamfiel}
 Thus due to the Frobenius
 theorem, locally they generate an integral submanifold $S$ through a point
 $x_0$ and it is always possible to introduce the local coordinates $x^\mu=
 (x^A, x^i)$, $A=1,...,d-\rank(p)$, $i=d-\rank(p)+1, ..., d$ in the
 neighbourhood of $x_0$ such that $S$ can be described by $x^i=const$ and
 $x^A$ are the coordinates on $S$.  The restriction of a Poisson bracket to
 the functions on a submanifold $S$ yields again a Poisson bracket. Thus
 $p|_S$ is a non degenerate Poisson structure on $S$, i.e. a symplectic
 structure.  As a result in these coordinates $x^\mu= (x^A, x^i)$, $p$ has
 the following form \beq p^{\mu\nu} = \left ( \begin{array}{ll} p^{AB} & 0 \\
 0 & 0 \end{array} \right ) \eeq{formP} If one wishes by doing an appropriate
 coordinate transformations in $x^A$ one can bring $p|_S$ to the
 canonical Darboux form.
 Thus one can conclude that the Poisson manifold is foliated by the
 symplectic leaves.  In a generic coordinate system there is locally complete
 set of independent Casimir functions $\{f_a(x)\}$ of $p$ which have
 vanishing Poisson bracket with any function from $C^\infty({\cal M})$. Then
  in this coordinates locally the symplectic leaf can be given by the
 conditions $f_a(x)=conts.$

Now let us apply this picture to our model with two Poisson tensors, $l$
and $m$.
Since the combinations $(l\pm m)$ are nondegenerate, the
 Poisson brackets, defined by the bi-vectors $l$ and $m$,
 can not have common  Casimir functions.
 Thus one can choose the following two sets of Casimir
 functions:  $\{ \phi_a(x), a=1,...,n_l \}$ which span the kernel of $l$ and
 $\{ \psi_i(x), i=1, ..., n_m\}$ which span the kernel of $m$. The
 conditions $\phi_a(x)=const$ define the symplectic submanifolds ${\cal S}_l$
 of dimension\footnote{Since the manifold ${\cal M}$ is
 even dimensional complex manifold the codimension of symplectic leaf is even and
 thus the number of Casimir functions is also even.}
 $(d-n_l)$ which are symplectic leaves of the Poisson structure
 $l$. The conditions $\psi_i(x)=const$ define the symplectic submanifolds
  ${\cal S}_m$ of dimension $(d-n_m)$ which are symplectic leaves of the
  Poisson structure $m$. Moreover the conditions $\phi_a(x)=const$ and
 $\psi_i(x)=const$ together define the submanifolds ${\cal S}_{lm}$ of the
 whole manifold ${\cal M}$
 \beq  {\cal S}_{lm} = {\cal S}_{l} \bigcap {\cal S}_{m} \subset {\cal M}
 \eeq{sssmmm}
 and ${\cal S}_{lm}$ is a submanifold
  of two symplectic manifolds, ${\cal S}_l$ and ${\cal S}_m$. Therefore the
 submanifolds ${\cal S}_{lm}$ can be described in terms of symplectic
 geometry. There are three numbers which would characterize ${\cal S}_{lm}$:
 the dimension ${\cal S}_{lm}$, the rank of restriction of the symplectic
 structure to ${\cal S}_{lm}$ from ${\cal S}_l$ and the rank of  restriction of the symplectic
 structure to ${\cal S}_{lm}$ from ${\cal S}_m$. Thus these three numbers together
 with the dimensions of the symplectic leaves ${\cal S}_l$ and ${\cal S}_m$ can be used
 for the description of the local geometry of the given type of the manifolds.
 The K\"ahler geometry is the extreme case of this geometry when, for example, the symplectic leaves 
 ${\cal S}_l$ coincide with the whole manifold ${\cal M}$ and ${\cal S}_m$ vanish.

The kernels of $l$ and $m$ are related in an obvious way to the  the (left)
kernels of $(J_+ + J_-)$ and $(J_- - J_+)$. In the work \cite{Ivanov:ec} it
 has been proved that the spaces orthogonal to the kernels of $(J_+\pm J_-)$
 and $[J_+, J_-] \equiv J_+J_- -J_-J_+$ (the product of $J_\pm$ is understood
 here as a standard matrix multiplication)
  are always integrable, and this proof was enough complicated.
  In
 the context of the Poisson geometry these facts become trivial as one can
 see from above discussion.  The spaces orthogonal to the kernels of
  $(J_+\pm J_-)$ correspond to the symplectic leaves of $l$ and $m$
  respectively.  The space, which is orthogonal
  to the kernel of $[J_+, J_-]$,
  corresponds to ${\cal S}_{lm}$.

To clarify the picture presented one can look at a special cases
(extreme in a sense) of this geometry.
 Let us start form the case when the complex structures commute ($[J_+, J_-]=0$).
 This case has been studied in the details in the original work \cite{Gates:1984nk}.
  The corresponding geometry is given by the local product of two symplectic manifolds.
 The symplectic leaf of $l$ ($m$) is the kernel of $m$ ($l$). The product of two complex
 structures gives rise to an integrable local product structure (i.e., $\Pi =J_+ J_-$,
  $\Pi^2=I$ and ${\cal N}^\rho_{\,\,\mu\nu}(\Pi)=0$). Hence the submanifolds projected by
 $1/2(I\pm \Pi)$ are symplectic leaves.

 Another interesting example corresponds to the  case when
 $[J_+, J_-]$ nondegenerate on $\cal M$.
 Thus $l$ and $m$ are invertiable and we can define their inverses
\beq
{\bf l}_{\mu\rho} l^{\rho\nu}=\delta^\mu_{\,\,\nu},\,\,\,\,\,\,\,\,\,\,\,\,\,\,\,
 {\bf m}_{\mu\rho} m^{\rho\nu} = \delta^{\mu}_{\,\,\nu} .
\eeq{defiinvertlm}
 The 2-forms ${\bf l}$ and ${\bf m}$ are closed
\beq
 d{\bf l} =0,\,\,\,\,\,\,\,\,\,\,\,\,\,\,\,\,
 d{\bf m} = 0
\eeq{twocloform}
 and the manifold ${\cal M}$ is bi-symplectic. In this case the torsion is expressible
 through the Nijenhuis tensor of $J_{\pm}$
\beq
 {\cal N}^\mu_{\,\,\nu\rho}(J_+, J_-) = 2 [J_+, J_-]^\mu_{\,\,\sigma} g^{\sigma\gamma}
 H_{\gamma\nu\rho}
\eeq{resultwtoNT}
 where $[J_+, J_-]$ is invertiable. Thus the torsion vanishes if and
 only if both complex structures are {\it simultaneously} integrable.
 In the case of zero torsion we have the bi-K\"ahler case and the
 Poisson structures $l$ and $m$ are compatible (i.e., $[l, m]_s=0$).
 In this case we have a continuous family of Poisson
 structures given by
 $p(t)= (t^1 l+ t^2 m)$ (where $t=(t^1,t^2) \in RP^1$ is
 any element of a real projective space) and smoothly interpolating
 between $l$ and $m$.

In this section we have discussed mostly the geometrical problems
 related to the structure of symplectic leaves of the
two Poisson brackets.
 A separate consideration is required for the geometry around the
 irregular points where the Poisson tensor(s) change their rank.

\section{N=(4,4) supersymmetric sigma model}
\label{s:generalizations}

In this section we discuss the general supersymetric sigma model with $N>2$. As was established
 in \cite{Gates:1984nk} there are interesting models with N=4. Below we show that these
 N=(4,4) sigma models  have a rich underlying Poisson geometry.

The dimensional analysis requires the following form of the
 additional supersymmetry transformations
\beq
  \delta^{(i)}(\epsilon) \Phi^\mu
 =\epsilon^+ D_+ \Phi^\nu J^{(i)\mu}_{+\nu}(\Phi)
  + \epsilon^- D_- \Phi^\nu J^{(i)\mu}_{-\nu}(\Phi) ,\,\,\,\,\,\,\,\,\,\,\,i=2,3,..., N .
\eeq{N4transf}
 We want the transformations (\ref{N4transf}) to be the symmetries of the action (\ref{actionB})
 and satisfy the on-shell algebra
\beq
 [\delta^i(\epsilon_1), \delta^j(\epsilon_2)]\Phi^\mu =2i\delta^{ij} \epsilon_1^+
 \epsilon_2^+ \d_\+ \Phi^\mu
 +  2i\delta^{ij} \epsilon_1^- \epsilon_2^- \d_=  \Phi^\mu
\eeq{N4algebra}
 where $\delta^1(\epsilon)$ is the manifest supersymmetry transformation given by (\ref{susyI}).

 The analysis goes along the lines presented in Section~\ref{s:susy}. The additional supersymmetry
 requires $J^{(i)}_{\pm}$ to be the complex structures. $J^{(i)}_{\pm}$ satisfy the conditions
 (\ref{bla}), (\ref{nablJH}) and (\ref{intergpl}), while (\ref{complstruct}) and
 (\ref{inegrabtroB}) are replaced by
 the more stringent conditions
\beq
 J^{(i)\mu}_{\pm\nu} J^{(j)\nu}_{\pm\rho} + J^{(j)\mu}_{\pm\nu} J^{(i)\nu}_{\pm\rho}
 = - 2 \delta^{ij} \delta^\mu_{\,\,\rho},
\eeq{su2algeb}
 \beq
 H_{\delta\nu\lambda}\delta^{ij} = J^{(i)\sigma}_{\pm\delta} J^{(j)\rho}_{\pm\nu} H_{\sigma\rho\lambda} +
 J^{(i)\sigma}_{\pm\lambda} J^{(j)\rho}_{\pm\delta} H_{\sigma\rho\nu}+
 J^{(i)\sigma}_{\pm\nu} J^{(j)\rho}_{\pm\lambda} H_{\sigma\rho\delta} .
\eeq{inegrabtroBN4}
 Thus for N=4 supersymmetry\footnote{N=3 supersymmetry implies N=4 and
 there is a bound on the number of supersymmetries, $N\leq 4$ (for the details see, e.g.
 \cite{Alvarez-Gaume:hm}).} the manifold has two quaternionic
  structures and therefore a dimensionality that is a multiple of four.
 Introducing the K\"ahler form $\omega_{\pm}^{(i)} = gJ_{\pm}^{(i)}$ the relations (\ref{definitdJ})
 and (\ref{HtermJplm}) are still valid. This description of N=(4,4) models has been presented in
 \cite{Gates:1984nk} (see also \cite{Gates:1994yk} and \cite{Gates:1995aj} for the recent 
 discussion of N=4 models). 

 Following the discussion of Section~\ref{s:Poisson} we can rewrite the above conditions in terms of
 the contravariant antisymmetric tensors $j^{(i)}_{\pm}= - J^{(i)}_{\pm}g^{-1}$.
 After the simple calculation we arrive at the algebra
\beq
 [j_{+}^{(i)}, j_{+}^{(j)} ]_s = - 2\delta^{ij} h ,\,\,\,\,\,\,\,\,\,\,\,\,\,\,\,\,
 [j_{-}^{(i)}, j_{-}^{(j)} ]_s =  2\delta^{ij} h ,\,\,\,\,\,\,\,\,\,\,\,\,\,\,\,\,
 [j_{+}^{(i)}, j_{-}^{(j)} ]_s =0
\eeq{alebrjN4}
As before we can introduce the linear combinations of $j^{(i)}_{\pm}$
\beq
 l^{(i)} = \frac{1}{2} (j_+^{(i)} + j_-^{(i)}),\,\,\,\,\,\,\,\,\,\,\,\,\,\,\,\,\,\,\,\,
 m^{(i)} = \frac{1}{2} (j_+^{(i)} - j_-^{(i)})
\eeq{redefN4}
 and rewrite the algebra (\ref{redefN4}) as follows
\beq
 [l^{(i)}, l^{(j)} ]_s = 0 ,\,\,\,\,\,\,\,\,\,\,\,\,\,\,\,\,
 [m^{(i)}, m^{(j)} ]_s =  0 ,\,\,\,\,\,\,\,\,\,\,\,\,\,\,\,\,
 [l^{(i)}, m^{(j)} ]_s = -\delta^{ij} h .
\eeq{algebN4lm}
 Thus there are six Poisson structures $l^{(i)}$ and $m^{(i)}$ such that  $l^{(i)}$ ($m^{(i)}$) is
 compatible with all other Poisson structures except $m^{(i)}$ ($l^{(i)}$).
 As in Section~\ref{s:Poisson} the algebra (\ref{algebN4lm}) gives the necessary and
  sufficient conditions for bi-quaternionic manifolds (the metric should be hermitian with
 respect to all complex structures) to admit the on-shell N=(4,4)
 supersymmetry algebra (\ref{N4algebra}).

The case when either $l^{(i)}$ or $m^{(i)}$ are zeros
 (i.e., either $J^{(i)}_+=-J^{(i)}_-$ or  $J^{(i)}_+=J^{(i)}_-$)
 corresponds to the hyperK\"ahler geometry. The hyperK\"ahler geometry corresponds
 to N=(4,4) supersymmetric sigma models without torsion, H=0.
 This case is automatically included in the previous
 discussion.

\section{Discussions}
\label{s:discussion}

In this short paper we argue that the appearance of the Poisson structures is a common
 feature of two dimensional sigma models with extended supersymmetry of type $(p,q)$
 with $2 \leq p$ and $2 \leq q$. We gave the explicit description of the \emph{most general}
  models with (2,2) and (4,4) supersymmetries. We briefly discussed the underlying geometry of
 the models. However, our discussion of the geometry is sketchy and we plan to come
 back to a detailed description of this type of geometry elsewhere.

 We believe that the new description of the general N=2 sigma model in terms of
 Poisson geometry opens new perspectives on old problems. Specifically we hope
 to study the twisted versions of the presented general N=2 sigma models.
 Another problem where the new description may be useful is the general description of
 N=2 boundary conditions (D-branes) of the corresponding sigma model \cite{Lindstrom:2002jb}.

{\bf Acknowledgements}:
We are grateful to Ulf Lindstr\"om for useful discussions and for
reading and commenting on the manuscript.
SLL is thankful to Lars Brink and Robert Marnelius for their very
warm hospitality at Chalmers University where he has begun this work
and to Seif Randjbar-Daemi at ICTP where the work was mostly
completed.  The work of SLL was supported in parts by the STINT
fellowship, INTAS grant 00-262, the RFBR grant 00-02-17956
and the grant E--00-33-184 from Russian Ministry of Education.

\appendix

\section{Appendix: (1,1) supersymmetry}

 In the appendix we list N=1 supersymmetric conventions we use mostly
 following \cite{Hitchin:1986ea}.

\noindent
We use  $(\+,=)$ as worldsheet indices, and $(+,-)$ as two-dimensional spinor
indices.  We also use superspace conventions where the pair of spinor
coordinates of the two-dimensional superspace are labelled $\th^{\pm}$,
and the covariant derivatives $D_\pm$ and supersymmetry generators
$Q_\pm$ satisfy
\beq
D^2_+ =i\d_\+ ,\,\,\,\,\,\,\,\,\,\,
D^2_- =i\d_=,\,\,\,\,\,\,\,\,\,\, \{D_+,D_-\}=0,\,\,\,\,\,\,\,\,\,\,
 Q_\pm = iD_\pm+ 2\th^{\pm}\d_{\pp}
\eeq{alg}
where $\d_{\pp}=\partial_0\pm\partial_1$.  In terms of the covariant
derivatives, a supersymmetry transformation of a superfield $\P$ is
then given by
\beq
\delta \P \equiv  i(\e^+Q_++\e^-Q_-)\P = -(\e^+D_++\e^-D_-)\P
 +2i(\e^+\th^+\d_\++\e^-\th^-\d_=)\P 
\eeq{tfs}


\begin{thebibliography}{6666}
%
\bibitem{Zumino:1979et}
B.~Zumino,
``Supersymmetry And Kahler Manifolds,''
Phys.\ Lett.\ B {\bf 87} (1979) 203.
%
\bibitem{Alvarez-Gaume:hm}
L.~Alvarez-Gaum\'e and D.~Z.~Freedman,
 ``Geometrical Structure And Ultraviolet Finiteness In The Supersymmetric Sigma Model,''
Commun.\ Math.\ Phys.\  {\bf 80} (1981) 443.
%
\bibitem{Gates:1983py}
S.~J.~Gates,
``Superspace Formulation Of New Nonlinear Sigma Models,''
Nucl.\ Phys.\ B {\bf 238} (1984) 349.
%
\bibitem{Gates:1984nk}
S.~J.~Gates, C.~M.~Hull and M.~Ro\v{c}ek,
``Twisted Multiplets And New Supersymmetric Nonlinear Sigma Models,''
Nucl.\ Phys.\ B {\bf 248} (1984) 157.
%
\bibitem{Hitchin:1986ea}
N.~J.~Hitchin, A.~Karlhede, U.~Lindstr\"om and M.~Ro\v{c}ek,
``Hyperkahler Metrics And Supersymmetry,''
Commun.\ Math.\ Phys.\  {\bf 108} (1987) 535.
%
\bibitem{Ikeda:1993fh}
N.~Ikeda,
``Two-dimensional gravity and nonlinear gauge theory,''
Annals Phys.\  {\bf 235} (1994) 435
[arXiv:hep-th/9312059].
%
\bibitem{SchStrobl}
P.~Schaller and T.~Strobl,
``Poisson structure induced (topological) field theories,''
Mod.\ Phys.\ Lett.\ A {\bf 9} (1994) 3129
[arXiv:hep-th/9405110].
%
\bibitem{CF1} 
A.~S.~Cattaneo and G.~Felder,
``A path integral approach to the Kontsevich quantization formula,''
Commun.\ Math.\ Phys.\  {\bf 212} (2000) 591
[arXiv:math.qa/9902090].
%
\bibitem{Kontsevich} 
M.~Kontsevich,
``Deformation quantization of Poisson manifolds, I,''
arXiv:q-alg/9709040.
%
\bibitem{CFT}
A.~S.~Cattaneo, G.~Felder and L.~Tomassini,
``From local to global deformation quantization of Poisson manifolds,''
arXiv:math.qa/0012228.
%
\bibitem{Fedosov} 
B.~V.~Fedosov, ``A Simple Geometrical Construction of Deformation Quantization,''
 J.\ Diff.\ Geom. {\bf 40} (1994) 213.
%
\bibitem{Schouten} 
J.~A.~Schouten, ``On differential operators
of first order in tensor calculus,'' in Convengo Int. Diff. Geom. Italia, 1953
Ed. Cremonese, Roma, 1954.
%
\bibitem{BV}
I.~A.~Batalin and G.~A.~Vilkovisky,
``Gauge Algebra And Quantization,''
Phys.\ Lett.\ B {\bf 102} (1981) 27.
%
\bibitem{Buscher:uw}
T.~Buscher, U.~Lindstr\"om and M.~Ro\v{c}ek,
``New Supersymmetric Sigma Models With Wess-Zumino Terms,''
Phys.\ Lett.\ B {\bf 202} (1988) 94.
%
%
\bibitem{Rocek:1991ze}
M.~Ro\v{c}ek,
``Modified Calabi-Yau manifolds with torsion,''
IASSNS-HEP-91-43
{\it Submitted to Proc. of Mirror Symmetry Workshop, MSRI, Berkeley, CA, May 1991}
%
%
\bibitem{Rocek:1991vk}
M.~Ro\v{c}ek, K.~Schoutens and A.~Sevrin,
``Off-shell WZW models in extended superspace,''
Phys.\ Lett.\ B {\bf 265} (1991) 303.
%
\bibitem{Ivanov:ec}
I.~T.~Ivanov, B.~b.~Kim and M.~Ro\v{c}ek,
``Complex Structures, Duality And WZW Models In Extended Superspace,''
Phys.\ Lett.\ B {\bf 343} (1995) 133
[arXiv:hep-th/9406063].
%
\bibitem{Sevrin:1996jr}
A.~Sevrin and J.~Troost,
``Off-shell formulation of N = 2 non-linear sigma-models,''
Nucl.\ Phys.\ B {\bf 492} (1997) 623
[arXiv:hep-th/9610102].
%
\bibitem{Bogaerts:1999jc}
J.~Bogaerts, A.~Sevrin, S.~van der Loo and S.~Van Gils,
``Properties of semi-chiral superfields,''
Nucl.\ Phys.\ B {\bf 562} (1999) 277
[arXiv:hep-th/9905141].
%
\bibitem{Spindel:1988nh}
P.~Spindel, A.~Sevrin, W.~Troost and A.~Van Proeyen,
``Complex Structures On Parallelized Group Manifolds And Supersymmetric Sigma Models,''
Phys.\ Lett.\ B {\bf 206} (1988) 71.
%
\bibitem{Spindel:1988sr}
P.~Spindel, A.~Sevrin, W.~Troost and A.~Van Proeyen,
``Extended Supersymmetric Sigma Models On Group Manifolds. 1. The Complex Structures,''
Nucl.\ Phys.\ B {\bf 308} (1988) 662.
%
\bibitem{Sevrin:1988ps}
A.~Sevrin, W.~Troost, A.~Van Proeyen and P.~Spindel,
``Extended Supersymmetric Sigma Models On Group Manifolds. 2. Current Algebras,''
Nucl.\ Phys.\ B {\bf 311} (1988) 465.
%
\bibitem{Yano1}
K.~Yano, ``Differential geometry on complex and almost
 complex spaces'' (Pergamon, Oxford, 1965)
%
\bibitem{vaisman}
I.~Vaisman, ``Lectures on the Geometry of Poisson Manifolds'',
 Progress in Mathematics, Vol 118
 (Birkh\"auser, Basel, 1994)
%
%
\bibitem{magri}
F.~Magri,
``A Simple Model Of The Integrable Hamiltonian Equation,''
J.\ Math.\ Phys.\  {\bf 19} (1978) 1156.
\bibitem{Gates:1994yk}
S.~J.~Gates,
``Why are there so many N=4 superstrings?,''
Phys.\ Lett.\ B {\bf 338} (1994) 31
[arXiv:hep-th/9410149].
%
\bibitem{Gates:1995aj}
S.~J.~Gates and S.~V.~Ketov,
``2D(4,4)l hypermultiplets. I: Diversity for N = 4 models,''
Phys.\ Lett.\ B {\bf 418} (1998) 111
[arXiv:hep-th/9504077].
%
\bibitem{Lindstrom:2002jb}
U.~Lindstr\"om and M.~Zabzine,
``N=2 Boundary conditions for non-linear sigma models and Landau-Ginzburg models,''
arXiv:hep-th/0209098.




\end{thebibliography}
\end{document}